\documentstyle[12pt,epsfig]{article}
\newcommand{\tdm}[1]{\mbox{\boldmath $#1$}}
\newcommand{\unit}[1]{\mbox{\rm #1}}
\begin{document}
\renewcommand{\textheight}{25cm}
\begin{titlepage}
\pagestyle{empty}
\vspace*{4cm}
\begin{center}
{\large\bf  Exclusive $\eta_c$ photo- and electroproduction at HERA \\
as a possible probe of the odderon singularity in QCD.}
\vspace{1.1cm}\\
{\sc J.~Czy\.zewski}$^{a,}$\footnote{ Fellow of the Polish 
             Science Foundation (FNP) scholarship for the year 1996},
         {\sc J.~Kwieci\'nski}$^b$, 
         {\sc L.~Motyka}$^a$, 
         {\sc M.~Sadzikowski}$^{b,1}$
\vspace{0.3cm}\\
$^a${\it Institute of Physics, Jagellonian University, 
Cracow, Poland}
\vspace{0.3cm}\\
$^b${\it Department of Theoretical Physics, \\
H.~Niewodnicza\'nski Institute of Nuclear Physics, 
Cracow, Poland}
\end{center}
\vspace{0.0cm}
\begin{abstract}  
Theory and phenomenology of the $\eta_c$ photo- and electroproduction is 
developed from the point of view of probing the odderon singularity in QCD 
which corresponds to the three gluon exchange  mechanism.  This mechanism 
leads to the cross-sections which are independent of $W^2$ for the exchange 
of three non-interacting gluons or exhibit increase with increasing  $W^2$ 
(or $1/x$) for the odderon intercept above unity. The $\eta_c$ 
electroproduction in the three gluon exchange mechanism is shown to be entirely 
controlled by the transversely polarised virtual photons. The magnitude of 
the $\eta_c$ photoproduction cross-section is estimated to be around 
11--45$\,$pb.  The $t$-dependence of the differential cross-section is 
also discussed.  
\end{abstract} 
\vspace{0.5cm}

\noindent
{\sf TPJU 23/96}\\
{\sf October 1996}
\end{titlepage}

The high-energy limit of perturbative QCD is at present fairly well 
understood \cite{LIPATOV1,LIPATOV2}.  Besides the pomeron 
[1--5]
one also expects presence of the so called ``odderon" singularity 
[1--3, 6, 7].
In the leading logarithmic 
approximation the pomeron corresponds to the exchange of two interacting 
(reggeized) gluons while the odderon is described by the three-gluon 
exchange.  Unlike pomeron which corresponds to the vacuum quantum numbers 
and so to  the positive charge conjugation 
the  ``odderon" is characterised by $C=-1$ (and I=0)  i.e.\ it carries 
the same quantum 
numbers as the $\omega$ Regge pole.  The (phenomenologically) determined 
intercept $\lambda_{\omega}$ of the $\omega$ Regge pole is 
approximately equal to $1/2$ 
\cite{DOLA}.  The novel feature of the odderon singularity corresponding 
to the gluonic degrees of freedom is the potentially very high value of its 
intercept $\lambda_{odd}\gg\lambda_{\omega}$.  The exchange of the 
three (noninteracting) gluons alone generates singularity with intercept 
equal to unity while interaction between gluons described in the leading 
logarithmic 
approximation by the BKP equation \cite{BARTELS,KP} is even capable to boost 
the odderon intercept above unity \cite{GLN}.  The energy dependence 
of the amplitudes corresponding to $C=-1$ exchange becomes similar to the 
diffractive ones which are controlled by the pomeron exchange.

It has been argued in \cite{MSN} that a very useful measurement which 
might test presence of the QCD odderon is the exclusive 
 photo (or electro) production 
of the $\eta_c$ meson at HERA.  Main merit of this process is that 
 the potential contribution  of the $\omega$ Reggeon  
to this process is expected to be strongly suppressed due to the Zweig rule. 
Moreover, presence of ``large" scale $m_c^2$ justifies the use of 
perturbative QCD.  The measurement of the 
$\eta_c$ photo or electroproduction acting as a useful tool    
for probing the QCD odderon is similar to the measurement of the $J/\Psi$ 
photo (or electro) production which probes the BFKL pomeron
implied by QCD \cite{RYSKIN,BRODSKY}.

\begin{figure}[h]
\begin{center}
\mbox{\epsfig{file=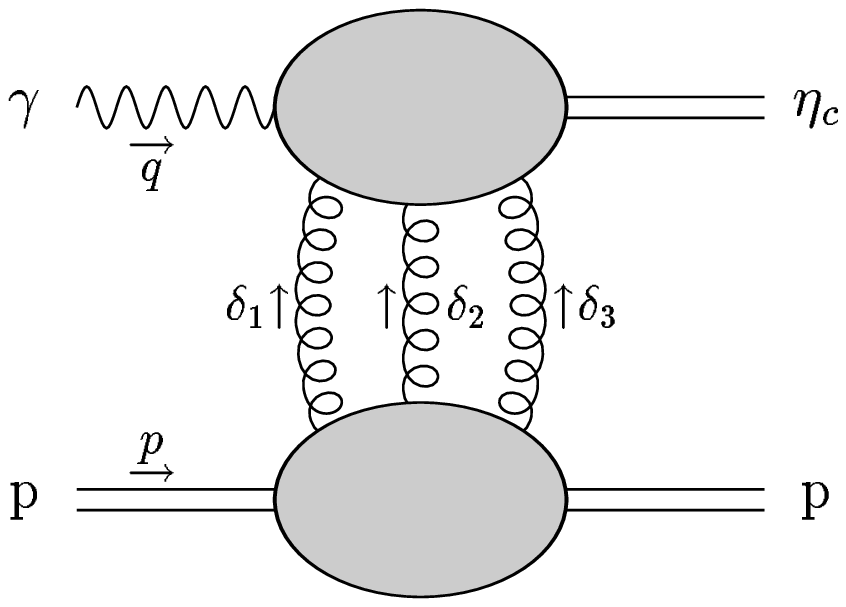,width=7cm}}
\vspace{-0.7cm}\\
{\small Fig.1: The kinematics of the three-gluon exchange of the 
process $\gamma^*  p \rightarrow \eta_c p$} 
\end{center}
\end{figure}

The main aim of our paper is to quantify  theoretical and phenomenological 
description of the $\eta_c$  photo- and electroproduction within
the three gluon 
exchange mechanism. Other processes which might probe the QCD
odderon 
in $\gamma^{*}p$ and in $\gamma \gamma$ interactions were
discussed in refs. \cite{GINZBURG,ZHITN}

The kinematics of the three gluon exchange diagram to the
process: 
$\gamma^{*}(q) + p \rightarrow \eta_c + p$  
 is illustrated in
Fig.~1.   It is convenient to introduce the light-like vectors $p^{\prime}$ 
and $q^{\prime}$:
$$  p^{\prime}=p+{M^2\over W^2}q$$
\begin{equation} 
q^{\prime}= p+xq, 
\label{llike} 
\end{equation} 
where  $x=Q^2/2pq$, $Q^2=-q^2$ and $W^2=(p+q)^2$. 
We assume that the four
momenta $p$ and $q$ are collinear and choose the frame where 
$p^{\prime}$ and $q^{\prime}$ have only ``$-$" and ``$+$" components 
respectively. We define the $\pm$ components of the four-vector $a^{\mu}$ as 
$a^{\pm}=
a^0 \pm a^{3}$. 
           
The amplitude for the process 
$\gamma^{*}(q) + p \rightarrow \eta_c + p$ can be written as below: 
\begin{equation}
A^{\mu}= {pq\over 16} {5\over 6} {1\over 3!}\int {d^2\tdm\delta_{t1}\over (2\pi)^4}  
{d^2\tdm\delta_{t2}\over (2\pi)^4} 
\Phi^{\mu}_{\gamma}
(Q^2,\tdm\delta_{t1}, \tdm\delta_{t2}, \tdm\Delta_t){1\over  
\tdm\delta_{t1}^2 \tdm\delta_{t2}^2 
\tdm\delta_{t3}^2} 
\Phi_p( \tdm\delta_{t1}, \tdm\delta_{t2}, \tdm\Delta_{t}), 
\label{a1}
\end{equation}
where the factor $5/6$ is the colour factor. $\Delta$ denotes 
the momentum transfer i.e.\ $\Delta=(p_{\eta_c}-q)$ 
and $\Delta= \delta_1+\delta_2+ \delta_3$ with all the momenta 
defined in Fig.~1.  
In the high-energy limit, $W^2 \rightarrow \infty$, we have 
$t=(q-p_{\eta_c})^2 =-\tdm\Delta_t^2$.   
  
The ``impact factors" $\Phi_i$ are related to the integrals of the 
the double  discontinuities of the amplitudes describing the upper 
and lower parts of the diagrams of Fig.~1.   To be precise, we
have: 
 
$$ 
\Phi_{\gamma}^{\mu}=\int ds_1 ds_{12} \unit{disc}_{s_1,s_{12}} 
{T^{\mu}_{\gamma+++}\over (q^{+})^3},$$

\begin{equation} 
\Phi_p=\int  ds_1^{\prime} ds_{12}^{\prime} 
\unit{disc}_{s_1^{\prime},s^{\prime}_{12}} {T_{p---}\over (p^{-})^3},
\label{impdisc} 
\end{equation} 
where: 
$$s_1=(q+\delta_1)^2,$$
$$s_{12}= (q+\delta_1+\delta_2)^2,$$
$$s^{\prime}_1=(p-\delta_1)^2,$$
\begin{equation}
s^{\prime}_{12}= (p-\delta_1-\delta_2)^2.
\label{ss} 
\end{equation} 
 The indices ``$+++$" or ``$---$"  correspond to the exchanged gluons.
    
The relevant diagrams describing 
the impact factor $\Phi^{\mu}_{\gamma}$ are given in Fig.~2.  
Besides the diagrams presented in Fig.~2 one has also to include
those with the reversed direction of the quark lines.                  
The calculation of the corresponding discontinuities is standard
i.e.\ for the diagram of Fig.~1a we have: 
\newpage
$$
\unit{disc}_{s_1,s_{12}} T^{\mu}_{\gamma+++}=  
e_c g_s^3(m_c^2) pq\int_0^1 dz d\alpha\int {d^2\tdm k\over (2\pi)^4}(2\pi)^3
{V_{\eta_c}[(2k-q-\Delta)^2]\over
[(k-\Delta)^2-m_c^2]} 
$$
$$ 
{\unit{Tr}\gamma^{\mu} [\gamma(q\!+\!k)\!+\!m_c] \gamma^5
[\gamma(k\!-\!\Delta)\!+\!m_c] \gamma^{+}
[\gamma(k\!-\!\delta_1\!-\!\delta_2)\!+\!m_c] \gamma^{+}
[\gamma(k\!-\!\delta_1)\!+\!m_c] \gamma^{+}
[\gamma k\!+\!m_c] \over (k^2-m_c^2)}
$$
\begin{equation}
\delta((q+k)^2-m_c^2)\,
\delta((\delta_1+\delta_2-k)^2-m_c^2)\,
\delta((\delta_1-k)^2-m_c^2)
\label{trasa}
\end{equation}
where $g_s(m_c^2)$ denotes the strong interaction coupling 
 evaluated at the scale $m_c^2$ and $e_c=q_ce$ with $q_c=2/3$.   
Similar expressions  describe contributions of 
the remaining 
diagrams of Fig.~2.   
The variables $z$ and $\alpha$ denote the corresponding Sudakov
parameters appearing in the decomposition of the 
momentum $k$ of the quark (antiquark) in terms of the  basic 
four momenta $p^{\prime}$ and $q^{\prime}$ i.e.:
\begin{equation}
k=-zq^{\prime}+\alpha p^{\prime} +k_t, 
\label{k}
\end{equation}
where the transverse four momentum $k_t$ is orthogonal to 
$p^{\prime}$ and to $q^{\prime}$.  
The vertex function $V_{\eta_c}$ together with the off-shell
propagator corresponding to the off-shell  quark (antiquark) 
adjacent to the $\eta_c$ meson can be related to the 
infinite-momentum-frame wave function of $\eta_c$. 
After evaluating the traces we find that the impact factor 
$\Phi_{\gamma}^{\mu}$ has only transverse components i.e.\ $\mu=i$, 
$i=1,2$ different from zero.

\begin{figure}[t]
\begin{center}
\mbox{\epsfig{file=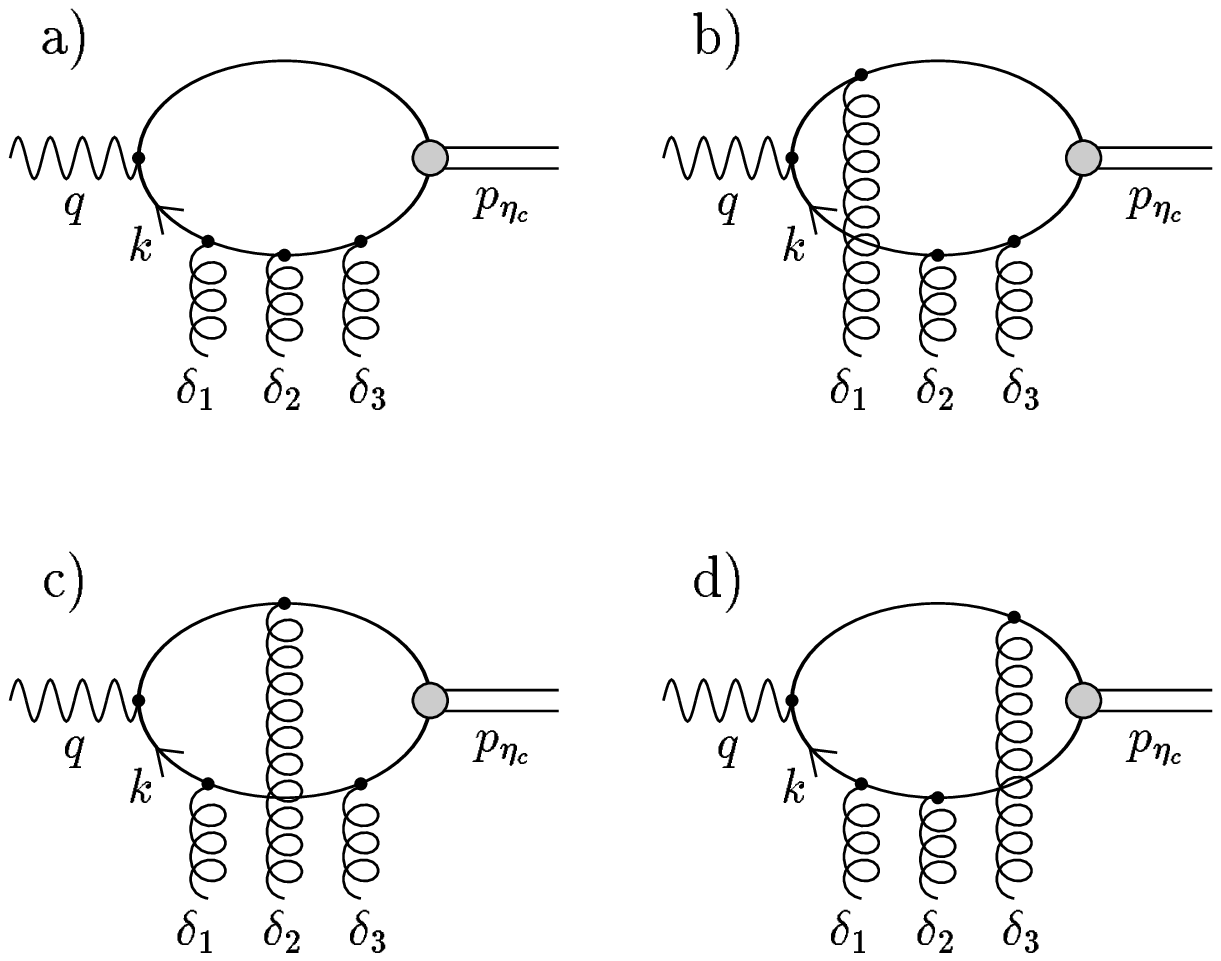,width=9cm}}
\vspace{-0.7cm}\\
{\small Fig.2:
The diagrams contributing to the impact factor
$\Phi_{\gamma}^{\mu}$ }
\end{center}
\end{figure}

When estimating the  integrals defining the impact
factor 
$\Phi_{\gamma}$ we assume that the dominant region of 
integration over $dz$ and
$d^2\tdm k$ is that which corresponds to the (off-shell) quark 
(antiquark) line at the $\eta_c \rightarrow c\bar c$ vertex 
being close to its mass-shell.  This gives $z\approx 1/2$ 
and $\tdm k_{ct}-\tdm k_{\bar ct} \approx \tdm\Delta_t$ where 
$\tdm k_{ct}$ and $\tdm k_{\bar c t}$ denote the transverse momenta 
of the charmed quark (antiquark) at the 
$\eta_c \rightarrow c\bar c$ vertex.      
The final expression for the impact factor then reads: 
\begin{equation}
\Phi^{i}_{\gamma}=
{16\over \pi}e_c g_s^3(m_c^2) {m_{\eta_c}\over 2}  \epsilon_{ij} 
{\Delta_{tj}\over \tdm\Delta_t^2} \left[
{\tdm\Delta_t^2\over Q^2\!+\!4m_{c}^2\!+\!\tdm\Delta_t^2}+
\sum_{k=1}^3{2\tdm\delta_{kt}\tdm\Delta_t-\tdm\Delta_t^2\over 
Q^2\!+\!4m_{c}^2\!+\!(2\tdm\delta_{kt}\!-\!\tdm\Delta_t)^2}\right] C,  
\label{ai}
\end{equation}  
where the constant $C$ can be determined by the  radiative width of the
meson $\eta_c$ 
\begin{equation}
C={16\pi^3\over 3 e_c^2} \sqrt{{\pi \Gamma_{\eta_c \rightarrow
\gamma \gamma} \over m_{\eta_c}}}. 
\label{gamma}
\end{equation}
When deriving the formula (\ref{gamma}) 
we have set $m_c=m_{\eta_c}/2$.  Similar approximation 
was 
adopted in the numerator of  formula (\ref{ai}) but we kept 
$m_c$ different from  $m_{\eta_c}/2$ in the denominators of  this formula. 
In our calculations we set $m_{\eta_c}=2.98\,$GeV, $m_c=1.4\,$GeV and  
$\Gamma_{\eta_c \rightarrow\gamma \gamma} = 7$keV \cite{ARMSTRONG}.
\vspace{-0.5cm}

\begin{figure}[h]
\begin{center}
\mbox{\epsfig{file=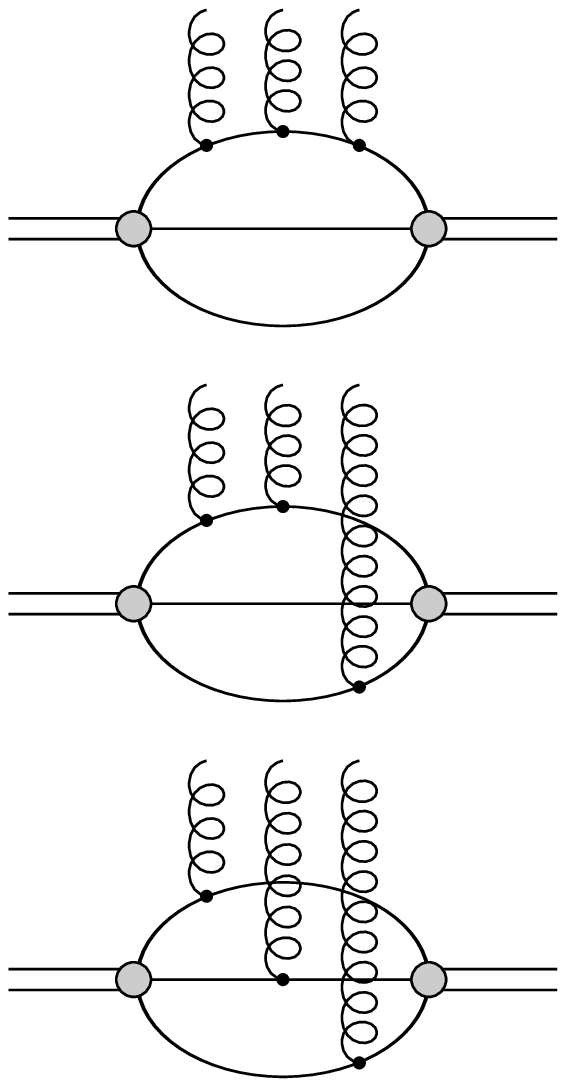,width=5cm}}
\vspace{-0.7cm}\\
{\small Fig.3:
The diagrams contributing to the impact factor $\Phi_p$ }
\end{center}
\end{figure}

The diagrams contributing to the 
 impact factor $\Phi_p$ are presented  in Fig.~3.  They give \cite{FK}: 
\begin{equation} 
\Phi_p=8(2\pi)^2\bar g^3 [F(\tdm\Delta_t,0,0)-\sum_i^3 
F(\tdm\delta_{it},\tdm\Delta_t-\tdm\delta_{it},0)
+2F(\tdm\delta_{1t},\tdm\delta_{2t},\tdm\delta_{3t})], 
\label{proton}
\end{equation}
where the form-factor $F$ will be assumed to have the form: 
\begin{equation}
F(\tdm\delta_{1t},\tdm\delta_{2t},\tdm\delta_{3t})=
{A^2\over A^2+ [(\tdm\delta_{1t}-\tdm\delta_{2t})^2 +
(\tdm\delta_{2t}-\tdm\delta_{3t})^2 +(\tdm\delta_{1t}-\tdm\delta_{3t})^2]/2}
\label{f}
\end{equation}
with $A \approx m_{\rho}/2$. We also set $\bar g^2/
(4\pi)=1$ which corresponds approximately to the magnitude of
the 
coupling which was used in the estimate of the hadronic
cross-sections within the two gluon exchange model \cite{GS}.

It should be noted that both impact factors vanish whenever 
$\tdm\delta_{ti}=0$ for $i=1$, $i=2$ or $i=3$.  This property 
of the impact factors which follows from the fact that the three
gluons couple to the colour singlets guarantees cancellation 
of  potential infrared singularities in the integral 
(\ref{a1})   at $ \tdm\delta_{ti}^2=0$.  The impact factor
$\Phi_{\gamma}^{\mu}$ also  
vanishes for $\tdm\Delta_t=0$. 

The photoproduction cross-section is related in a standard way
to the 
amplitude $A^i$ at $Q^2=0$ 
\begin{equation} 
{d\sigma\over dt}(\gamma + p \rightarrow \eta_c + p)= 
{1\over 16 \pi W^4} {1\over 2} \sum_{i=1}^2 |A^i|^2.  
\label{photo}
\end{equation} 
The electroproduction cross-section is given by the following
formula: 
\begin{equation}
{d\sigma\over dt dQ^2dy}(e+p \rightarrow e^{\prime}+ \eta_c + p)= 
{\alpha\over \pi y Q^2} (1-y+{y^2\over 2}) 
{d\sigma\over dt}(\gamma^* + p \rightarrow \eta_c + p), 
\label{electro}
\end{equation} 
where 
\begin{equation}
y={qp\over p_ep}
\label{y}
\end{equation} 
with $p_e$ denoting the four momentum of the incident electron.  The virtual 
photoproduction cross-section is given by the same formula as (\ref{photo}) 
but with $A^i$ calculated for $Q^2$ different from $0$.
 
\begin{figure}[h]
\begin{center}
\mbox{\epsfig{file=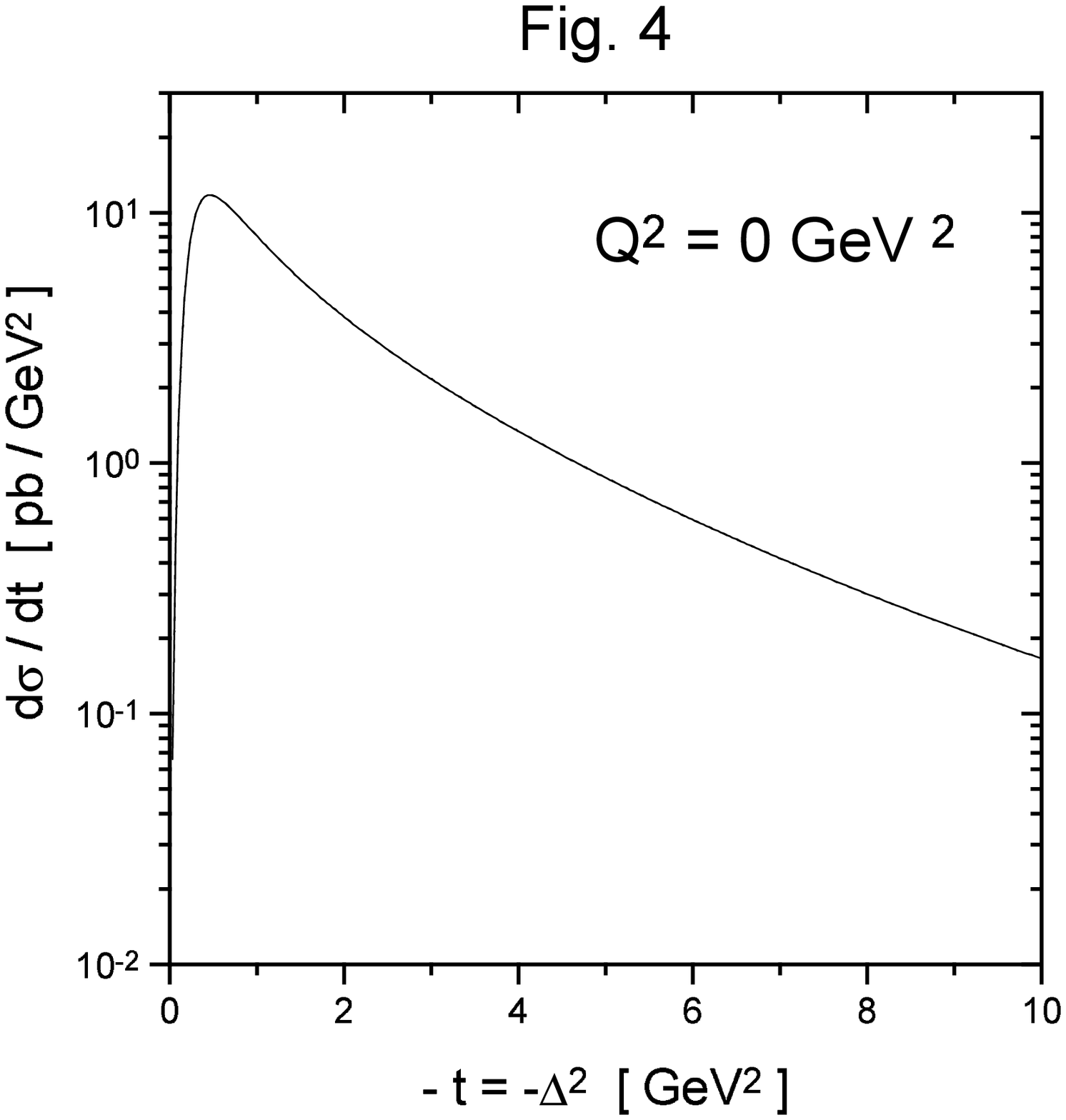,width=9cm}}
\vspace{-0.2cm}\\
\parbox{14cm}
{\small Fig.4:
The differential cross-section for the process 
$\gamma p \rightarrow \eta_c p$ plotted as the function of the
squared momentum transer $t$ }
\end{center}
\end{figure}

In Fig.~4 we show    the photoproduction
cross-section plotted as the function of $t$. The cross-section vanishes at 
$t=0$ and has a
maximum at $t\approx  -0.5\,$GeV$^{2}$ where it reaches the value 
$\sim  11\,$pb$\,$GeV$^{-2}$.   The photoproduction cross-section is  
also characterized by the relatively weak $t$ dependence at large $t$ where 
the diagrams with the gluons coupling to three different quarks
in a proton dominate.  It should be noted that the two gluon 
exchange mechanism of diffractive $J/\Psi$ production generates 
strong suppression of the differential cross-section in the
large-$t$ region due to the nucleon form-factor effects.  This fact can
be useful in experimental separation of the $\eta_c$  mesons  
produced through the odderon exchange from those which are 
the decay products 
of the diffractively produced $J/\Psi$-s.  The photoproduction 
cross-section can be fitted to the following simple form: 
\begin{equation}
{d\sigma\over dt}(\gamma + p \rightarrow \eta_c + p)= a{|t|^3\over 
[(b+t^2)(c+|t|^{3/2})]^2}, 
\label{fit}
\end{equation}
where $a=4.4\,$nb$\,$GeV$^6$, $b=0.075\,$GeV$^4$ and $c=21.4\,$GeV$^3$.  
The total 
 cross-section for the process $\gamma + p \rightarrow \eta_c + p$  
 is estimated to be equal to 11$\,$pb.
 
   In Fig.~5 we plot the virtual photoproduction cross-section 
for $Q^2=25\,$GeV$^2$.  
This cross-section is much smaller than the real
photoproduction cross-section.

The cross-sections presented in Fig.~4 and Fig.~5 were calculated 
assuming the exchange of  three elementary gluons.  This
mechanism leads to 
the 
odderon intercept $\lambda_{odd}$ equal to unity and so the corresponding
cross-sections 
are independent of $W^2$.    Interaction between gluons can
boost the odderon intercept above unity and the variational
estimate gives the following bound on $\lambda_{odd}$ \cite{GLN}: 
\begin{equation}
\lambda_{odd} -1 \ge 0.13(\lambda_{pom}-1),
\label{lodderon} 
\end{equation} 
\begin{figure}[t]
\begin{center}
\mbox{\epsfig{file=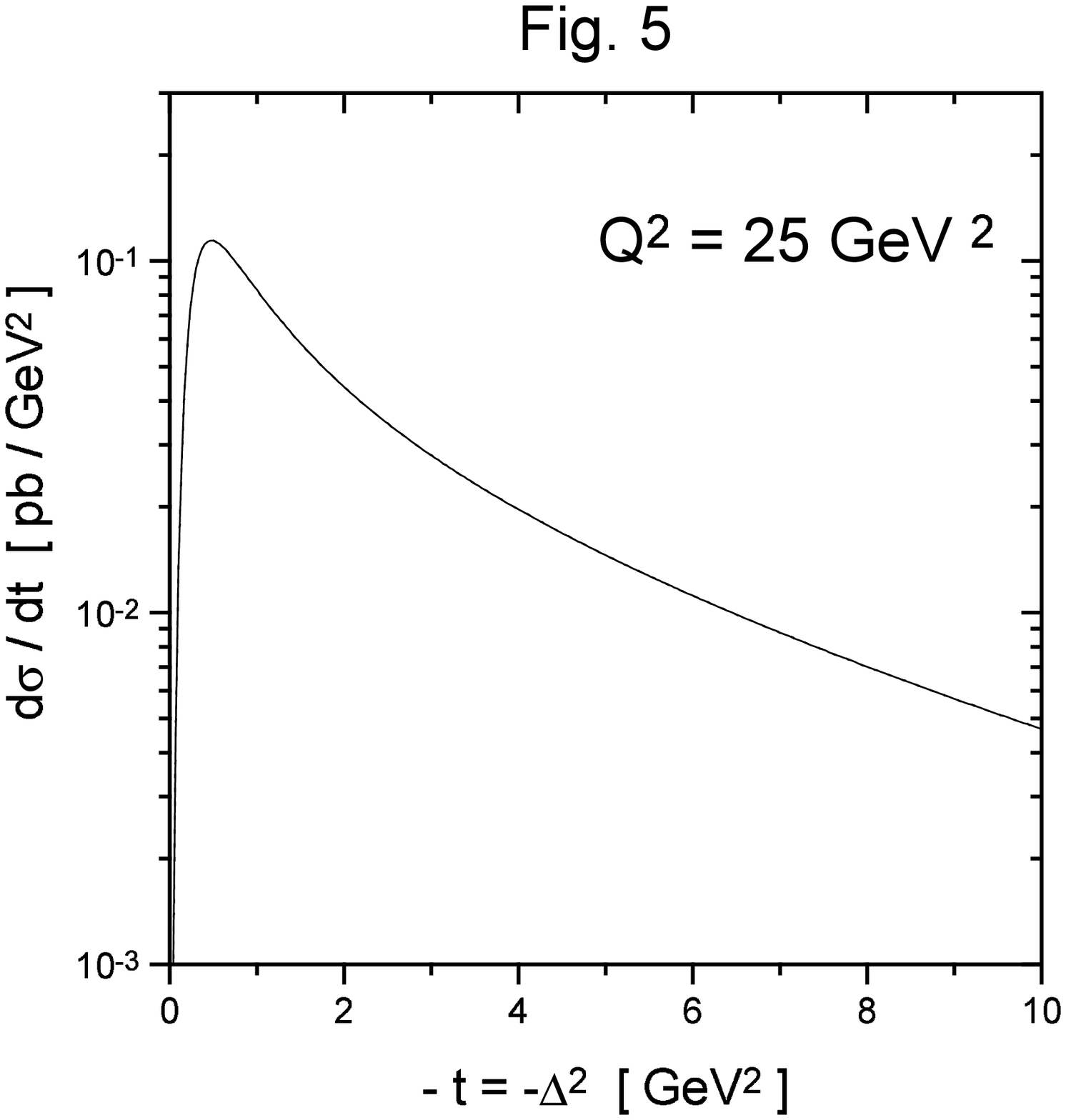,width=9cm}}
\vspace{-0.2cm}\\
\parbox{14cm}
{\small Fig.5:
The differential cross-section for the process 
$\gamma^*  p \rightarrow \eta_c p$ plotted as the function of 
the squared momentum transfer $t$.  The magnitude of the virtuality $Q^2$
of the virtual photon $\gamma^*$ was set equal to $25\,$GeV$^2$  }
\end{center}
\end{figure}
where $\lambda_{pom}$ denotes the intercept of the BFKL pomeron \cite{BFKL}. 
For the typical value of the BFKL pomeron  $\lambda_{pom} = 1.5$ we get 
$\lambda_{odd} \ge 1.07$.  
We may approximately  accomodate the possibility that 
$\lambda_{odd}>1$ by multiplying the cross-sections
(\ref{photo}, 
\ref{electro}) by the
$W^2$ dependent enhancement factor $D$:
\begin{equation} 
D=\bar x^{2 (\lambda_{odd}-1)}, 
\label{d}
\end{equation}
where 
\begin{equation}
\bar x={m_{\eta_c}^2+Q^2\over W^2+Q^2}. 
\label{xbar}
\end{equation}
Assuming that $\lambda_{odd} \sim 1.1$ we get $D \sim 4$ in
the 
HERA kinematical range i.e.\ the cross-sections presented in Figs~4 
and 5 can be larger by  a factor equal to 4 or so.  They
should 
also exhibit weak increase with increasing $W^2$ given by
(\ref{d}).     

To summarize, we have discussed the cross-sections for the $\eta_c$
photo- and electroproduction in the high-energy limit as possible 
probes of the QCD odderon.    The
magnitude of the total photoproduction cross-section was
estimated to be equal about 11--45$\,$pb.  This cross-section should be
independent of 
$W^2$ or exhibit weak increase with increasing $W^2$ ($ \sim
(W^2)^{0.2}$ or so).  The $\eta_c$ photoproduction is not the
only process which may probe the QCD odderon and the photoproduction 
of other $c \bar c$  bound states characterised by positive
charge conjugation should presumably be useful as well.
Theoretical 
analysis of these and related processes is  in progress \cite{CKMS}. 

\section*{Acknowledgments} 
We thank Halina Abramowicz, Krzysztof Golec-Biernat, Genya Levin and  
J.E.~Olsson for illuminating discussions.  
We would like also to acknowledge a very useful comment from 
Dima Ivanov and Lech Szymanowski concerning our results.
One of us (MS) thanks the Theoretical Physics Laboratory in DESY
for warm hospitality.  This research was partially supported by
the Polish 
State Committee for Scientific Research (KBN) grants 
Nos.:~2P~03B~083~08, 2P~03B~231~08, 2P~302~076~07 and by the Volkswagen 
Stiftung.
JC was also supported by the Polish-German Collaboration Foundation grant
FWPN no.~1441/LN/94.

\newpage

\end{document}